\documentclass[aps,prl,twocolumn,superscriptaddress,showpacs,amsmath,amssymb]{revtex4-1}
\usepackage{graphicx,bbm}
\usepackage{hyperref}
\usepackage{mathtools}
\usepackage{color}
\usepackage{braket}

\renewcommand{\d}[1]{\ensuremath{\operatorname{d}\!{#1}}}

\newcommand{\tr}[1]{\ensuremath{\operatorname{tr}\!{#1}}}
\newcommand{\jhalf}{{\textstyle\frac{\delta}{2}}}
\newcommand{\nhalf}{{\textstyle\frac{N-1}{2}}}
\newcommand{\pauli}{\boldsymbol\sigma}

\begin{document}

\title{Integrable Trotterization: Local Conservation Laws and Boundary Driving}
\author{Matthieu Vanicat}
\author{Lenart Zadnik}
\author{Toma\v{z} Prosen}
\affiliation{Faculty of Mathematics and Physics, University of Ljubljana, Jadranska 19, SI-1000 Ljubljana, Slovenia}

\begin{abstract}
We discuss a general procedure to construct an integrable real--time trotterization of interacting lattice models. As an illustrative example we consider a spin-$1/2$ chain, with continuous time dynamics described by the isotropic ($XXX$) Heisenberg Hamiltonian. For periodic boundary conditions local conservation laws are derived from an inhomogeneous transfer matrix and a boost operator is constructed. In the continuous time limit these local charges reduce to the known integrals of motion of the Heisenberg chain. 
In a simple Kraus representation we also examine the nonequilibrium setting, where our integrable cellular automaton is driven by stochastic processes at the boundaries.
We show explicitly, how an exact nonequilibrium steady state density matrix can be written in terms of a staggered matrix product ansatz. 
This simple trotterization scheme, in particular in the open system framework, could prove to be a useful tool for experimental simulations of the lattice models in terms of trapped ion and atom optics setups.
\end{abstract}

\maketitle

{\bf Introduction.--} Quantum integrable systems out of equilibrium are a topic of intense current research, both theoretically and experimentally \cite{JSTAT2016}.
Universal relaxation properties of integrable systems based on the hypothesis of local equilibrium, given by the generalized Gibbs ensemble \cite{GGE}, depend crucially on the knowledge of local and quasi-local charges of the system. Furthermore, the study of nonequilibrium quantum transport problem in integrable systems has been fruitful in the context of the boundary driven Lindblad equation \cite{pro2015}, where the properties of
the nonequilibrium steady state can be connected to a rigorous existence of ballistic transport at high temperature.
However, all these concepts have, so far, been developed for autonomous, time-independent systems, while much recent interest also goes in the direction of periodically-driven (Floquet) many-body systems, in particular in the connection to topological phases \cite{FloquetPhases} and time crystals \cite{TimeCrystals}. 

Periodically time-dependent system can be naturally viewed as a Trotter approximation of a one-dimensional interacting, continuous time model, on which state-of-the-art matrix product simulation methods are based \cite{vidal}. This {\em trotterized evolution} is itself a discrete time dynamical system -- in fact a reversible quantum cellular automaton \cite{SchuWerner}. The {\em quantum transfer matrix} approach, proposed by Kl\"umper {\em et al}, \cite{klumper,klumper2,klumper4}, presents a way of generating Bethe-ansatz integrable systems of this form and of improving the efficiency of transfer-matrix renormalization group calculations \cite{klumper3}. Independently, this kind of integrable quantum evolution has been used in lattice discretizations of continuous field theories such as the famous sine-Gordon model \cite{faddeev,devega,volkov}.

In this Letter we discuss an integrable unitary circuit which may be viewed as a Trotter formulation of some integrable continuous Hamiltonian, or an integrable Floquet (periodically driven) quantum chain. We take the simplest and physically perhaps most relevant example of the {\em isotropic Heisenberg spin-$1/2$ model}, i.e., the $XXX$ model. First we define the dynamical system from the inhomogeneous (staggered) transfer matrix. We then proceed to show how two independent families of local conservation laws can be constructed along with a boost operator. Finally we present an integrable steady state density matrix of a boundary driven dissipative protocol which is formulated in terms of Krauss maps acting on the boundary spins. This is the first time, an explicit solution of an interacting discrete time quantum Markov chain has been presented.

{\bf Integrable trotterization.--}
Consider a chain of $N$ spins-$1/2$. Instead of a continuous time evolution of the density matrix given by the {\em Liouville--von Neumann} equation
\begin{equation}
i\,\frac{\d{\rho_t}}{\d{t}} = [H,\rho_t]\label{mat_continuous},
\end{equation}
and generated by a hamiltonian $H$, we would like to construct a discrete--time map 
\begin{equation}
\rho_{t+1} = \mathcal{U}\,\rho_t\,\mathcal{U}^{\dagger},
\end{equation}
$\mathcal{U}$ being a unitary propagator, so that  in the appropriate limit the original continuous dynamics \eqref{mat_continuous} is recovered. The propagator $\mathcal{U}$ should be expressed in terms of a finite sequence of operators $U$ acting locally on the spin chain. Moreover it should commute with some extensive family of local operators, $\{Q_k\}$, generated by a transfer matrix, so that we can declare the model to be integrable.
From a mathematical point of view the advantage of such trotterization would be, that the expectation values $\bra{\psi}\mathcal{U}^n\ket{\psi}$
could be interpreted as equilibrium partition functions of some vertex models, for which a lot of computational techniques have already been developed.

As an example we consider a specific nontrivial SU(2)-invariant spin-$1/2$ model, although the construction should easily be generalizable to other Yang-Baxter integrable models.
For the local propagator $U$, acting on any two neighboring spins of the chain, take the $\check{R}$-matrix of the $XXX$ model \cite{faddeevLH},
\begin{align}
U_{j,j+1}=\check{R}_{j,j+1}(\delta),\qquad\check{R}(\lambda)=\frac{1+i\,\lambda\,P}{1+i\,\lambda}\label{len_RU}
\end{align}
Indices denote the spins acted upon by the operator, while $P$ denotes a permutation. If we denote Pauli matrices by ${\boldsymbol\sigma}=(\sigma^x,\sigma^y,\sigma^z)$ and identity by $1$, the latter can be written as $P_{j,j+1}=\frac{1}{2}\,\big(1+{\boldsymbol \sigma}_j\cdot{\boldsymbol \sigma}_{j+1}\big)$. Real parameter $\delta$ can be interpreted as a spin coupling constant and is crucial in recovering the continuous dynamics. The particular choice of the 
normalization of $\check{R}$ ensures unitarity. $U$ chosen in this way, appears as one of the simplest unitary operations on a pair of spins-$1/2$ one can think of -- 
it swaps the spins with a probability amplitude proportional to the coupling parameter $\delta$. It can be interpreted as an elementary quantum gate.

\begin{figure}[h]
\includegraphics[width=200pt]{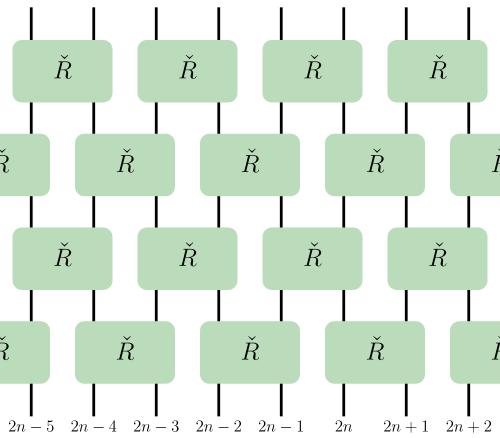}
\caption{Time evolution of the model. Each time step consists of two half-steps. In the first one, we apply gates $U_{2j-2,2j-1}$, and in the second one, $U_{2j-1,2j}$. The protocol thus shifts for one site each half-step.}
\label{fig:propagation}
\end{figure}

Suppose now, that $N$ is an even integer. The full unitary propagator acting on the whole chain of $N$ spins-$1/2$ can be defined as
\begin{align}
\mathcal{U}=\mathcal{U}_{even}\,\mathcal{U}_{odd}=\prod_{j=1}^{N/2}U_{2j-1,2j}\prod_{k=1}^{N/2}U_{2k,2k+1}.\label{len_propagator}
\end{align}
We have imposed periodic boundary conditions so that the sites $N+1$ and $1$ are equivalent. The full propagator $\mathcal{U}$ can be interpreted as a two--step discrete--time Floquet dynamics or a quantum cellular automaton, see figure \ref{fig:propagation}.
The first step is carried by the action of $\mathcal{U}_{odd}$ which updates all even--odd numbered spins, while the second step $\mathcal{U}_{even}$ updates, in the same manner, spins on odd--even sites.

From the construction of $\mathcal{U}$ it is easy to realize that in the infinitesimal coupling limit, $\delta=-J\Delta t$, with $\Delta t=t/n$ and $n$ being very large, we get
$U_{1,2}\sim1-i\,\Delta t\, h_{12}$, where $h_{12}=J P_{12}$ is the local hamilonian density of the $XXX$ spin-$1/2$ chain and
\begin{align}
\lim_{n\to \infty}\mathcal{U}^n=\exp{(-i\,t\,H)}\label{trotter}
\end{align}
As expected, this is just the Trotter formula for the Hamiltonian of the $XXX$ spin-$1/2$ model
\begin{align}
H=\frac{J}{2}\,\sum_{j=1}^N\big(1+{\boldsymbol \sigma}_{j}\cdot{\boldsymbol \sigma}_{j+1}\big).
\end{align}
A similar trotterization scheme, using the {\em quantum transfer matrix} method has been used for computing quantum dynamics for a particular subset of initial states \cite{Piroli}, or to simplify computation of dynamical correlation functions \cite{klumper4}. It also corresponds to the row-transfer matrix Floquet integrability as defined in \cite{Gritsev}.

Since we have taken an $\check{R}$-matrix for the local propagator \eqref{len_RU} the integrability of this trotterization scheme is not surprising. It stems from a family of inhomogeneous (staggered) {\em transfer operators} $T(\lambda)$ which are in involution, $[T(\lambda),T(\mu)]=0$, for all values of the {\em spectral parameters} $\lambda$ and $\mu$. This involution is described in Appendix A. The inhomogeneous transfer operators act on the whole spin chain and can be expressed as
\begin{align}
T(\lambda)=\tr{_0 \left( \prod^{\longleftarrow}_{1\le j \le N} R_{0,j}(\lambda-(-1)^j\jhalf)\right)},\label{len_T}
\end{align}
where we have denoted $R(\lambda)=P \check{R}(\lambda)$. The trace is taken over the {\em auxiliary space}, which is a copy of a spin-$1/2$ space, used to facilitate the matrix--product formalism. Elementary evaluation of the transfer operator at two particular points, $\delta/2$ and $-\delta/2$, now yields the following expression for the unitary
propagator (\ref{len_propagator})
\begin{align}
\mathcal{U}=\left[T(-\jhalf)\right]^{-1}\,T(\jhalf).\label{len_U}
\end{align}
In particular this implies integrability of this model.

In lattice discretizations of continuous field theories, see e.g. \cite{faddeev,volkov,devega}, the time evolution can generically be put into such perspective. One example is the  sine-Gordon field theory and its discretized version -- the quantum Hirota equation \cite{faddeev}. It is a trotterization of the quantum Volterra model \cite{volkov}. The entire hierarchy of conservation laws, which we are going to present now, can also be constructed in that case \cite{progress}.

{\bf Local integrals of motion.--} The transfer operators \eqref{len_T} generate local integrals of motion through the logarithmic differentiation. Due to the inhomogeneity we have two families of such local conservation laws, given by evaluating the logarithmic derivatives at the points $\delta/2$ and $-\delta/2$ respectively,
\begin{align}
Q_n^+=\frac{\d{^n}}{\d{\lambda^n}}\log{T(\lambda)}\Bigr|_{\lambda=\substack{\jhalf}},\; Q_n^-=\frac{\d{^n}}{\d{\lambda^n}}\log{T(\lambda)}\Bigr|_{\lambda=\substack{-\jhalf}}.\label{len_Q}
 \end{align}
Conservation laws are invariant under translation for {\em two} lattice sites.
Explicit computation shows that the local terms of $Q_n^\pm$ act nontrivially on $2n+1$ neighbouring sites. Terms of $Q_n^-$ are, however,  shifted for one lattice site to the right, with respect to those of $Q_n^+$. Explicitly we can write, for example,
\begin{align}
Q^+_1=\sum_{n=1}^{N/2}q^{[1,+]}_{2n-2,2n-1,2n},\quad
Q^-_1=\sum_{n=1}^{N/2}q^{[1,-]}_{2n-1,2n,2n+1},\label{len_Q1}
\end{align}
with the three--site local densities being 
\begin{align}
q^{[1,\pm]}_{1,2,3}&=\frac{i}{2(1+\delta^2)}\big[{\boldsymbol\sigma}_1\cdot{\boldsymbol\sigma}_2+\,{\boldsymbol\sigma}_2\cdot{\boldsymbol\sigma}_3+\delta^2\,{\boldsymbol\sigma}_1\cdot{\boldsymbol\sigma}_3\mp  \notag\\[1em]
&\mp \delta\,{\boldsymbol\sigma}_1\cdot\big({\boldsymbol\sigma}_2\times{\boldsymbol\sigma}_3\big)
\big].
\end{align}
We have subtracted the trivial terms proportional to the identity. We have also computed the densities of $Q_2^\pm$, which are written in Appendix C (see Supplementary material \cite{suppmat}). At this point we would like to stress, that in the continuous--time limit, $\delta\to 0$, both of the derivatives \eqref{len_Q} become the same and converge to the standard charges of $XXX$ model \cite{grabowski}. In particular $\lim_{\delta\to 0} Q_1^\pm \propto H$, i.e., the first pair of charges become the hamiltonian of the $XXX$ model.

Both sets of conservation laws, $Q^+_n$ and $Q^-_n$, can now be equipped with the {\em boost operation} -- a ladder mapping which transforms lower--order conservation laws into higher--order ones and thus facilitates their computation. In our case it takes the form
\begin{align}
&[B,Q_n^\pm]=Q_{n+1}^\pm, \label{len_BQ}\\
&B=\sum_{n=1}^{N/2}\Big(n\!\!\!\mod \frac{N}{2}\Big)\,\mathbbm{R}'_{2n-3,2n-2|2n-1,2n}(0),\label{len_B}
\end{align}
$B$ being the {\em boost operator}. In equation \eqref{len_B} $f'(0)$ denotes the derivative with respect to $\lambda$ at the point $\lambda=0$, while
$\mathbbm{R}_{12|34}(\lambda)=\check{R}_{23}(\lambda-\delta)\check{R}_{12}(\lambda)\check{R}_{34}(\lambda)\check{R}_{23}(\lambda+\delta)$ is the {\em four--point R-matrix}. The factor $n$ has been 
considered modulo $N/2$ due to periodic boundary conditions. The local terms of the boost operator can also be explicitly written in terms of three-site $SU(2)$ invariant product of vectors of Pauli matrices
\begin{align}
\begin{aligned}
\mathbbm{R}'_{12|34}(0)=\frac{i}{2(1+\delta^2)}\big[{\boldsymbol\sigma}_1\cdot{\boldsymbol\sigma}_2+{\boldsymbol\sigma}_3\cdot{\boldsymbol\sigma}_4+\\[1em]
+2\,{\boldsymbol\sigma}_2\cdot{\boldsymbol\sigma}_3+\delta^2\,{\boldsymbol\sigma}_2\cdot{\boldsymbol\sigma}_4+\delta^2\,{\boldsymbol\sigma}_1\cdot{\boldsymbol\sigma}_3+\\[1em]
+\delta\,{\boldsymbol\sigma}_1\cdot\big({\boldsymbol\sigma}_2\times{\boldsymbol\sigma}_3\big)
-\delta\,{\boldsymbol\sigma}_2\cdot\big({\boldsymbol\sigma}_3\times{\boldsymbol\sigma}_4\big)
\big],
\end{aligned}
\end{align}
where the identity component has been subtracted as it does not affect the boost procedure. Again, note, that in the limit $\delta\to 0$, boost operator becomes the first moment of the $XXX$ model's hamiltonian, in accordance with \cite{grabowski}. For the derivation of the boost relation, which is similar as in the case of a homogeneous transfer operator \cite{grabowski}, see Appendix B of the supplementary material \cite{suppmat}. 

{\bf Dissipative boundaries.--}
We now wish to study the behavior of the model when a dissipative protocol is performed at the boundaries. 
The goal is to write a discrete time version of the Lindblad equation with local 
dissipators located at the first and the last site of an open chain and to find the steady state of the protocol that it defines. For convenience, the length $N$ of the chain is now taken to be an {\em odd} integer. The dissipative boundaries can be modelled by two pairs of Kraus operators
\begin{equation}
 K_0 = \frac{\mathbbm{1}+\sigma_1^z}{2} +\sqrt{1-\gamma_L}\,\, \frac{\mathbbm{1}-\sigma_1^z}{2}, \quad K_1=\sqrt{\gamma_L}\,\sigma_1^{+}
\end{equation}
for the left boundary, and 
\begin{align}
 \overline{K}_0 = \frac{\mathbbm{1}-\sigma_N^z}{2} + \sqrt{1-\gamma_R}\,\,\frac{\mathbbm{1}+\sigma_N^z}{2}, \quad \overline{K}_1 =\sqrt{\gamma_R}\,\sigma_N^{-}
\end{align}
for the right one. Note, that they satisfy the trace preservation conditions, 
$\sum_{j=0}^1 K_j^{\dagger} K_j = \mathbbm{1}$, $\sum_{j=0}^1 \overline{K}_j^{\dagger} \overline{K}_j = \mathbbm{1}$.
We write the dynamics of the density matrix as a two--step discrete--time protocol
\begin{align}
 \rho_{t+1} = \hat{\mathcal{M}} \,\rho_t, \qquad \hat{\mathcal{M}}=\hat{\mathcal{M}}_{odd}\,\hat{\mathcal{M}}_{even},
\end{align}
where $\hat{\mathcal{M}}_{even}$ and $\hat{\mathcal{M}}_{odd}$ are two {\em completely--positive maps} defined as 
\begin{align}
\begin{aligned}
 \hat{\mathcal{M}}_{even}\,\rho & = \sum_{j=0}^1 \overline{K}_j\, \mathcal{U}_{even}\, \rho \,\mathcal{U}_{even}^{\dagger} \,\overline{K}_j^{\dagger}\\
 \hat{\mathcal{M}}_{odd}\,\rho & = \sum_{j=0}^1 K_j\, \mathcal{U}_{odd}\, \rho \,\mathcal{U}_{odd}^{\dagger}\, K_j^{\dagger}
\end{aligned}
\end{align}
The unitary propagators $\mathcal{U}_{even}$ and $\mathcal{U}_{odd}$ are defined similarly as in the periodic case, but this time with additional boundary magnetic fields, represented in terms of unitary matrices $B_1 = \exp(ib_L \sigma_1^z)$ and $\overline{B}_N = \exp(ib_R \sigma_N^z)$,
\begin{align}
 \mathcal{U}_{even} = \prod_{j=1}^{\nhalf}U_{2j-1,2j} \overline{B}_N, \quad \mathcal{U}_{odd} =  B_1 \prod_{j=1}^{\nhalf}U_{2j,2j+1}
\end{align}
Parameters $b_L$ and $b_R$ correspond to strengths of magnetic fields at the left and the right edge of the chain, respectively.
This process has a very natural interpretation in terms of a repeated interaction protocol \cite{Karevski,Attal} where, periodically, in each half time step, left-most/right-most spin is brought into interaction with a
fresh up/down polarized spin. As such, the protocol has probably a more straightforward experimental implementation than the corresponding boundary driven Lindblad equation \cite{pro2015,Prosen11b} 
(see also \cite{Popkov13}) which is obtained in the continuum limit $\delta\to 0$.

As is shown in Appendix D of the supplementary material \cite{suppmat}, the stationary problem $\rho_{\infty} = \hat{\mathcal{M}}\,\rho_{\infty}$ can be solved exactly for the {\em stationary state} $\rho_\infty$. We will see, that the structure of the stationary state remarkably resembles that of the inhomogeneous transfer matrix introduced in the previous section. Let us state the ansatz and the result. The stationary state is searched for in the Cholesky form as
\begin{align}
\rho_{\infty} = \frac{\Omega^{\dagger}\Omega}{\tr(\Omega^{\dagger}\Omega)},\label{mat_cholesky}
\end{align}
where $\Omega$ is a triangular matrix built in an inhomogeneous (staggered) matrix--product form
\begin{align}
&\Omega  =  D^{\otimes N}\bra{0}L_{a,1}(\lambda,s)L_{a,2}(\lambda-\delta,s)\cdots \nonumber \\
&\quad\qquad\qquad\cdots L_{a,N-1}(\lambda-\delta,s) L_{a,N}(\lambda,s)\ket{0}.\label{mat_omega}
\end{align}
By $D$ we have denoted a diagonal matrix which acts on the spin-$1/2$ space and depends on some arbitrary real  parameter $\chi$, while $L(\lambda,s)$ is the well known {\em Lax matrix} of the isotropic spin-$1/2$ Heisenberg model,
\begin{align}
 D =\begin{bmatrix}
           \chi^{1/4} & 0 \\
           0 & \chi^{-1/4}
          \end{bmatrix},\quad
         L(\lambda,s) = \begin{bmatrix}
           i\lambda+S^{z} & S^{-} \\ S^{+} & i\lambda - S^{z}
          \end{bmatrix}.\label{mat_L}
\end{align}
In the matrix--product expression \eqref{mat_omega}, Lax matrices are equipped with two indices. The first index denotes the position of the spin in the chain, upon which $L(\lambda,s)$ acts, while the letter $a$ denotes the auxiliary space. The latter is now an infinite-dimensional space with basis $\{\ket{k}\}_{k=0}^{\infty}$, upon which operators $S^{z}$ and $S^\pm$ act. These operators satisfy $sl_2$ algebraic relations $[S^{z},S^\pm]=\pm S^\pm$ and $[S^{+},S^{-}]=2S^{z}$ and can take the following explicit form
\begin{align}
\begin{aligned}
 S^{z} & = \sum_{k=0}^{\infty} (s-k) \ket{k}\bra{k} \\
 S^{+} & = \sum_{k=0}^{\infty} (k+1) \ket{k}\bra{k+1} \\
 S^{-} & = \sum_{k=0}^{\infty} (2s-k) \ket{k+1}\bra{k}.
\end{aligned}
\end{align}
For $s$ being an integer  or a half-integer, we recover the standard $2s+1$ dimensional unitary spin-$s$ representation from these relations.  However, in our case $s$ and the other two parameters, $\lambda$ and $\chi$, are complex and real numbers set by the boundary conditions,
\begin{align}
 \lambda & = \frac{\delta}{2}\left(\frac{1}{1-e^{-2ib_R}\sqrt{1-\gamma_R}}-\frac{e^{2ib_L}\sqrt{1-\gamma_L}}{1-e^{2ib_L}\sqrt{1-\gamma_L}} \right) \notag\\[1em]
 s & = \frac{i\delta}{2}\left(\frac{1}{1-e^{-2ib_R}\sqrt{1-\gamma_R}}+\frac{e^{2ib_L}\sqrt{1-\gamma_L}}{1-e^{2ib_L}\sqrt{1-\gamma_L}} \right) \notag\\[1em]
 \chi & =  \frac{\gamma_L}{\gamma_R}\frac{2\big[1-\cos(2b_R)\sqrt{1-\gamma_R}\big]-\gamma_R}{2\big[1-\cos(2b_L)\sqrt{1-\gamma_L}\big]-\gamma_L}.
\end{align}
For derivation of this exact solution, see Appendix D of the supplementary material \cite{suppmat}.

Note that the stationary state shares a lot of common features with the solution of the similar continuous--time model, i.e., the Lindblad--driven $XXX$ spin-$1/2$ chain (the most general one presented in the review \cite{pro2015}). An important difference to our discrete-time case is the introduction of the inhomogeneous (staggered) structure of the matrix--product ansatz. This is needed in order to deal with the two--step discrete--time dynamics. Conceptually similar procedure is used to write the steady state and Markov eigenvectors of the classical, boundary driven Bobenko cellular automaton -- see \cite{pro2017}.

{\bf Conclusion.--} We have discussed a method of constructing an integrable real-time trotterization scheme of interacting lattice models, with both, the time evolution and the local conservation laws generated by an inhomogeneous transfer matrix. The discussion related to the hierarchy of local conservation laws can be understood in a wider context -- it relies solely on the existence of a unitary solution $R(\lambda)$ of the Yang-Baxter equation, which generates the cellular automaton through the inhomogeneous transfer matrix. The extension of our construction to quasi-local conservation laws should be straightforward following the procedures discussed in the review \cite{Ilievski16}.

Time evolution of our system is a sequence of local quantum gates and can be understood either as a Floquet driven system or a quantum cellular automaton.  As a particular example we have studied, in detail, a trotterized $XXX$ spin-$1/2$ model. For the periodic lattice, we have considered local integrals of motion and derived the boost operator for the inhomogeneous transfer operator. In the continuous time limit these conservation laws become those of the $XXX$ model. Apart from that, we have exactly solved the system, which is described by the trotterization scheme in the bulk, while being coupled with a thermal reservoir at the boundaries.

From the point of view of a field theorist, these results provide a method to treat dissipative boundary conditions in a light--cone discretization, while for an experimentalist, the described discrete--time quantum protocol could become a simple paradigm of a spin chain simulator realized by subsequent application of SWAP-like quantum gates, corresponding to the local propagator and additional boundary dissipative processes \cite{swap,ion}.

{\bf Acknowledgements.--} LZ would like to thank Katja Klobas and Marko Medenjak for useful comments on the manuscript. We thank Lorenzo Piroli and Fabian Essler for fruitful discussions.
The work has been supported by Advanced Grant 694544 -- OMNES of the European Research Council (ERC) and by grants N1-0025, N1--0055 of the Slovenian Research Agency (ARRS).

\clearpage
\setcounter{equation}{0}
\section{Supplementary material}

{\bf Appendix A. Integrability structure.--} Our local propagator is equivalent to the so-called rational $\check{R}$-matrix of the six-vertex model, $U_{i,j}=\check{R}_{i,j}(\delta)$, where the spectral parameter has taken the value of the spin coupling parameter. This $\check{R}$-matrix satisfies the {\em braiding} relation
\begin{align}
\check{R}_{12}(\lambda)\check{R}_{23}(\lambda+\mu)\check{R}_{12}(\mu)=\check{R}_{23}(\mu)\check{R}_{12}(\lambda+\mu)\check{R}_{23}(\lambda).\label{len_RRR}
\end{align}
Together with $R$-matrix $R(\lambda)=P\check{R}(\lambda)$, which is the building block of our transfer operator, Eq.~(7) of the main text, it satisfies also the {\em intertwining relation}
\begin{align}
\check{R}_{23}(\lambda-\mu)R_{12}(\lambda)R_{13}(\mu)=R_{12}(\mu)R_{13}(\lambda)\check{R}_{23}(\lambda-\mu).\label{len_RLL}
\end{align}
Due to the symmetry of the $\check{R}$-matrix, $\check{R}_{23}=\check{R}_{32}$, the indices $2$ and $3$ in this relation can be exchanged between the un--checked $R$-matrices. Then it is straightforward to show $[T(\lambda),T(\mu)]=0$ for two arbitrary complex spectral parameters $\lambda$ and $\mu$.

{\bf Appendix B. Boost operator.--} In order to derive the boost operator, $B$, we rewrite the transfer matrix, Eq.~(7) of the main text, in terms of {\em double $L$-matrices}, defined as 
\begin{align} \mathbbm{L}_{0|12}(\lambda)=R_{0,2}(\lambda-\jhalf)R_{0,1}(\lambda+\jhalf).
\end{align} Obviously, $T(\lambda)=\tr{_0 \big(\mathbbm{L}_{0|N-1,N}(\lambda)\cdots \mathbbm{L}_{0|3,4}(\lambda)\mathbbm{L}_{0|1,2}(\lambda)\big)}$. The double $L$-matrix satisfies another intertwining relation,
\begin{align}
\begin{aligned}
&\mathbbm{R}_{12|34}(\lambda-\mu)\mathbbm{L}_{0|34}(\lambda)\mathbbm{L}_{0|12}(\mu)=\\
&=\mathbbm{L}_{0|34}(\mu)\mathbbm{L}_{0|12}(\lambda)\mathbbm{R}_{12|34}(\lambda-\mu),\label{len_RRLLLL}
\end{aligned}
\end{align}
where $\mathbbm{R}_{12|34}(\lambda)=\check{R}_{23}(\lambda-\delta)\check{R}_{12}(\lambda)\check{R}_{34}(\lambda)\check{R}_{23}(\lambda+\delta)$. Differentiating \eqref{len_RRLLLL} with respect to $\lambda$ at $\lambda=\mu$, one gets the so-called {\em Sutherland's relation}
\begin{align}
\begin{aligned}
&[\mathbbm{R}'_{12|34}(0),\mathbbm{L}_{0|34}(\lambda)\mathbbm{L}_{0|12}(\lambda)]=\\
&=\mathbbm{L}_{0|34}(\lambda)\mathbbm{L}'_{0|12}(\lambda)-\mathbbm{L}'_{0|34}(\lambda)\mathbbm{L}_{0|12}(\lambda).\label{len_sutherland}
\end{aligned}
\end{align}
In terms of the latter, one can derive the following result
\begin{align}
\begin{aligned}
&\big[\sum_{n=1}^{N/2}n\, \mathbbm{R}'_{2n-3,2n-2|2n-1,2n}(0),T(\lambda)\big]=\\
&=T'(\lambda)-\frac{N}{2}\tr{_0\big(\mathbbm{L}'_{0|N-1,N}(\lambda)...\mathbbm{L}_{0|1,2}(\lambda)\big)},\label{len_boost1}
\end{aligned}
\end{align}
which becomes simply the boost equation
\begin{align}
[B,T(\lambda)]=\frac{\d{}}{\d{\lambda}}T(\lambda),\label{len_boost2}
\end{align}
if we equate $\frac{N}{2}+1\equiv 1 \mod N/2$. The boost relation, Eq.~(12) of the main text, immediately follows. 

At this point a remark is necessary. Explicit calculation shows, that on the right hand side of Eq.~\eqref{len_boost1}, the identity components of the first and the second term cancel out exactly, while the left hand side contains no term proportional to an identity since it is traceless. Hence, when taking the equation {\em modulo} $N/2$, one has to take care, to remove not just the second term $-\frac{N}{2}\tr{_0\big(\mathbbm{L}'_{0|N-1,N}(\lambda)...\mathbbm{L}_{0|1,2}(\lambda)\big)}$, but also the identity component of $T'(\lambda)$ (which also contains prefactor $N/2\equiv 0\mod N/2$). To reparaphrase, the boost equation \eqref{len_boost2} should contain no terms proportional to an identity.

{\bf Appendix C: The second logarithmic derivatives.--} Here we write the densities of the second logarithmic derivatives $Q_2^\pm$. We have
\begin{align}
\begin{aligned}
Q^+_2=\sum_{n=1}^{N/2}q^{[2,+]}_{2n-2,2n-1,2n,2n+1,2n+2},\\[1em]
Q^-_2=\sum_{n=1}^{N/2}q^{[2,-]}_{2n-1,2n,2n+1,2n+2,2n+3},\label{len_Q2}
\end{aligned}
\end{align}
with the local five--site densities
\begin{align}
\begin{aligned}
&q^{[2,\pm]}_{1,2,3,4,5}=\frac{i}{2(1+\delta^2)^2}[\mp 2\delta\,\pauli_3\cdot\pauli_4\mp2\delta\,\pauli_4\cdot\pauli_5\\[1em]
&\pm2\delta\,\pauli_3\cdot\pauli_5-(1-\delta^2)\,\pauli_3\cdot(\pauli_4\times\pauli_5)-\pauli_2\cdot(\pauli_3\times\pauli_4)\\[1em]
&-\delta^2\,\pauli_2\cdot(\pauli_3\times\pauli_5)-\delta^2\,\pauli_1\cdot(\pauli_3\times\pauli_4)\\[1em]
&-\delta^4\,\pauli_1\cdot(\pauli_3\times\pauli_5)\pm \delta\,\pauli_2\cdot(\pauli_3\times\pauli_4\times\pauli_5)\\[1em]
&\pm \delta\,\pauli_1\cdot(\pauli_2\times\pauli_3\times\pauli_4)\pm \delta^3\,\pauli_1\cdot(\pauli_3\times\pauli_4\times\pauli_5)\\[1em]
&\pm \delta^3\,\pauli_1\cdot(\pauli_2\times\pauli_3\times\pauli_5)-\delta^2\,\pauli_1\cdot(\pauli_2\times\pauli_3\times\pauli_4\times\pauli_5)].
\end{aligned}
\end{align}
Once again, terms in $Q_2^-$ are shifted for one site to the right with respect to those of $Q_2^+$.

{\bf Appendix D: Proof of the stationarity of the density matrix.--}
The Lax operator, Eq.~(22) of the main text, along with the local propagator, Eq.~(3) of the main text,  which is itself an $\check{R}$-matrix, satisfies the RLL relation
\begin{align}
\begin{aligned}
 &\check{R}_{1,2}(\lambda-\mu)L_{a,1}(\lambda,s)L_{a,2}(\mu,s) = \\
 &=L_{a,1}(\mu,s)L_{a,2}(\lambda,s) \check{R}_{1,2}(\lambda-\mu).
\end{aligned}
\end{align}
Since the local propagator is unitary, it is easy to deduce from this relation, that the following two equations hold, 
\begin{align}
\begin{aligned}
&U_{1,2}\,L_{a,1}(\lambda,s)L_{a,2}(\lambda-\delta,s)=\\
&=L_{a,1}(\lambda-\delta,s)L_{a,2}(\lambda,s)\,U_{1,2},\\[1em]
&U_{1,2}\,L_{a,1}^{T_1}(-\lambda^*,s^*)L_{a,2}^{T_2}(-\lambda^*+\delta,s^*)=\\
&=L_{a,1}^{T_1}(-\lambda^*+\delta,s^*)L_{a,2}^{T_2}(-\lambda^*,s^*)\,U_{1,2}.\label{eq:exchange_LLt}
\end{aligned}
\end{align}
By $(\bullet)^{T_{1(2)}}$ we have denoted the transposition on the first (second) spin-$1/2$ space and by $(\bullet)^*$ the complex conjugation.

Another well--known fact is, that the $\check{R}$-matrix of the $XXX$ spin-$1/2$ chain commutes
with any tensor square $A\otimes A$ of a two by two matrix $A$. In particular, we have $U \, D \otimes D = D \otimes D \, U$ on the tensor product of two spin-$1/2$ spaces. Using these facts and
\begin{align}
\begin{aligned}
\Omega^\dagger=\bra{0}L_{a,1}^{T_1}(-\lambda^*,s^*)L_{a,2}^{T_2}(-\lambda^*+\delta,s^*)\cdots\times\\
\times\cdots L_{a,N}^{T_N}(-\lambda^*,s^*)\ket{0}D^{\otimes N}
\end{aligned}
\end{align}
along with ansatz, Eq.~(20) of the main text, we can deduce that the following relations should be satisfied on the boundaries,
\begin{align} 
&L_{a,N}^{T_N}(-\lambda^*+\delta,s^*) D_N^2L_{b,N}(\lambda-\delta,s)\ket{0}_{a}\ket{0}_{b} =\notag\\
&=\sum_{m=0}^1 \overline{K}_m \overline{B}_N L_{a,N}^{T_N}(-\lambda^*,s^*)D_N^2\times\notag\\
&\qquad\times L_{b,N}(\lambda,s)\overline{B}_N^{\dagger} \overline{K}_m^{\dagger}
\ket{0}_{a}\ket{0}_{b}
\label{eq:exchange_right_boundary}
\end{align}
and 
\begin{align}
&\bra{0}_{a}\bra{0}_{b}L_{a,1}^{T_1}(-\lambda^*,s^*)D_1^2L_{b,1}(\lambda,s)=\notag\\
&=\bra{0}_{a}\bra{0}_{b} \sum_{m=0}^1 K_m B_1 L_{a,1}^{T_1}(-\lambda^*+\delta,s^*)D_1^2\times\notag\\
&\qquad\times L_{b,1}(\lambda-\delta,s)B_1^{\dagger} K_m^{\dagger},
\label{eq:exchange_left_boundary}
\end{align}
in order for $\rho_{\infty} = \hat{\mathcal{M}}\,\rho_{\infty}$ to hold.

Now it is straightforward to realize that the values of parameters $\lambda$, $\chi$ and $s$, given by Eq. (24) of the main text, are necessary to fulfill the boundary relations \eqref{eq:exchange_right_boundary} and \eqref{eq:exchange_left_boundary}. The mechanism of this proof can be decomposed into two steps (consistently with the two--step dynamics). First $\hat{\mathcal{M}}_{even}\,\rho_{\infty}=\rho_{\infty}'$, where $\rho_{\infty}'$ has the same expression as $\rho_{\infty}$ but with $L(\lambda,s)$-s and $L(\lambda-\delta,s)$-s interchanged. In the second step we demand $\hat{\mathcal{M}}_{odd}\,\rho_{\infty}'=\rho_{\infty}$.


\begin{thebibliography}{99}

\bibitem{JSTAT2016}
P.~Calabrese, F.~H.~L.~Essler and G.~Mussardo, {\em Special issue on quantum integrability in out of equilibrium systems}, J. Stat. Mech. 064001 (2016), and articles therein.

\bibitem{GGE}
M.~Rigol, V.~Dunjko,~V. Yurovsky and M.~Olshanii, 
{\em Relaxation in a completely integrable manybody quantum system: An ab initio study of the dynamics of the highly excited states of 1D lattice
hard-core bosons}, Phys. Rev. Lett. \textbf{98}, 050405 (2007).

\bibitem{pro2015}
T.~Prosen,
{\em Matrix product solutions of boundary driven quantum chains}, J. Phys. A: Math. Theor. \textbf{48}, 373001  (2015).

\bibitem{FloquetPhases}
J.~Cayssol, B.~Dora, F.~Simon and R.~Moessner,
{\em  Floquet topological insulators}, Phys. Status Solidi RRL \textbf{7}, 101 (2013).

\bibitem{TimeCrystals}
F.~Wilczek, {\em Quantum Time Crystals}, Phys. Rev. Lett. \textbf{109}, 160401 (2012).

\bibitem{vidal}
G.~Vidal,
{\em Efficient Classical Simulation of Slightly Entangled Quantum Computations}, Phys. Rev. Lett. \textbf{91}, 147902 (2003).

\bibitem{SchuWerner} 
B.~Schumacher, R.~F.~Werner, {\em Reversible quantum cellular automata}, {\tt arXiv:quant-ph/0405174}.

\bibitem{klumper}
A.~Kl\"{u}mper, 
{\em Thermodynamics of the anisotropic spin-$1/2$ Heisenberg chain and related quantum chains}, Z. Phys. B \textbf{91}, 507 (1993).

\bibitem{klumper2}
A.~Kl\"{u}mper, 
{\em Integrability of quantum chains: Theory and applications to the spin-$1/2$ XXZ chain}, Lect. Notes in Phys. \textbf{645}, 349-379  (2004).

\bibitem{klumper4}
F.~G\"ohmann, M.~Karbach, A.~Kl\"umper, K.~K.~Kozlowski, J.~Suzuki,
{\em Thermal form-factor approach to dynamical correlation functions of
integrable lattice models}, J. Stat. Mech. {\bf 2017}, P113106 (2017).

\bibitem{klumper3}
J.~Sirker, A.~Kl\"umper,
{\em Temperature driven crossover phenomena in the correlation lengths of the
one-dimensional t-J model}, Euro. Phys. Lett \textbf{60}, 262 (2002).

\bibitem{faddeev}
L.~D.~Faddeev, A.~Yu.~Volkov,
{\em Hirota equation as an example of integrable symplectic map}, Letters in Mathematical Physics \textbf{32}, 125-135 (1994).

\bibitem{volkov}
A.~Yu.~Volkov,
{\em Quantum Volterra model}, Physics Letters A \textbf{167}, 345-355  (1992).

\bibitem{devega}
C.~Destri, H.~J.~De Vega,
{\em Light-cone lattice approach to fermionic theories in 2D: the massive Thirring model}, Nucl. Phys. B \textbf{290}, 363  (1987).

\bibitem{faddeevLH}
L.~D.~Faddeev, 
{\em How Algebraic Bethe Ansatz works for integrable model}, {\tt arXiv:hep-th/9605187} (1996).

\bibitem{Piroli} L.~Piroli, B.~Pozsgay and E.~Vernier, {\em From the quantum transfer matrix to the quench action: the Loschmidt echo in XXZ Heisenberg spin chains}, 
J. Stat. Mech. {\bf 2017}, 023106 (2017).

\bibitem{Gritsev}
V.~Gritsev, A.~Polkovnikov, {\em Integrable Floquet dynamics}, SciPost Phys. {\bf 2}, 021 (2017).

\bibitem{progress}
M.~Vanicat, L.~Zadnik, T.~Prosen,
{\em to be submitted.}

\bibitem{suppmat}
{\em Supplementary material associated with this manuscript.} 

\bibitem{grabowski}
M.~P.~Grabowski, P.~Mathieu
{\em Structure of the conservation laws in quantum integrable spin chains with short range interactions}, Ann. Phys. \textbf{243}, 299 (1995).

\bibitem{Karevski} D.~Karevski, T.~Platini, {\em Quantum nonequilibrium steady states induced by repeated interactions}, Phys. Rev. Lett. {\bf 102}, 207207 (2009).

\bibitem{Attal} S.~Attal, Y.~Pautrat, {\em From repeated to continuous quantum interactions}, Annales Henri Poincar\' e {\bf 7}, 59 (2006).

\bibitem{Prosen11b} T.~Prosen, {\em Exact Nonequilibrium Steady State of a Strongly Driven Open XXZ Chain}, Phys. Rev. Lett. {\bf 107}, 137201 (2011).

\bibitem{Popkov13} D.~Karevski, V.~Popkov, G.~M.~Sch\" utz, {\em Exact Matrix Product Solution for the Boundary-Driven Lindblad XXZ Chain}, Phys. Rev. Lett. {\bf 110}, 047201 (2013).

\bibitem{pro2017}
T.~Prosen, B.~Bu\v ca, {\em Exact matrix product decay modes of a boundary driven cellular automaton}, J. Phys. A: Math. Theor. \textbf{50}, 395002 (2017).

\bibitem{Ilievski16} E.~Ilievski, M.~Medenjak, T.~Prosen, L.~Zadnik, {\em Quasilocal charges in integrable lattice systems}, J. Stat. Mech. {\bf 2016}, 064008 (2016).

\bibitem{swap}
T. R. Tan, J. P. Gaebler, Y. Lin, Y. Wan, R. Bowler, D. Leibfried, D. J. Wineland, {\em Multi-element logic gates for trapped-ion qubits}, Nature {\bf 528}, 380 (2015).

\bibitem{ion}
J.~T.~Barreiro, M.~M\" uller, P.~Schindler, D.~Nigg, T.~Monz, M.~Chwalla, M.~Hennrich, C.~F.~Roos, P.~Zoller, R.~Blatt,  
{\em An open-system quantum simulator with trapped ions}, Nature {\bf 470}, 486 (2011).



\end{thebibliography}
\end{document}